\documentclass[10pt,letterpaper]{article}
\usepackage{opex3}
\usepackage{amsmath}
\usepackage{cite}
\usepackage{float}

\begin{document}

\title{Focusing Light Through Random Photonic Media By Binary Amplitude Modulation}

\author{D. Akbulut, T. J. Huisman, E. G. van Putten, W. L. Vos and A. P. Mosk}

\address{Complex Photonic Systems, Faculty of Science and Technology
and MESA+ Institute for Nanotechnology, University of Twente,
P.O.Box 217, 7500 AE Enschede, The Netherlands}

\email{d.akbulut@tnw.utwente.nl} 

\homepage{http://wavefrontshaping.com} 


\begin{abstract}
We study the focusing of light through random photonic materials using wavefront shaping. We explore a novel approach namely binary amplitude modulation. To this end, the light incident to a random photonic medium is spatially divided into a number of segments. We identify the segments that give rise to fields that are out of phase with the total field at the intended focus and assign these a zero amplitude, whereas the remaining segments maintain their original amplitude. Using 812 independently controlled segments of light, we find the intensity at the target to be 75$\pm$6 times enhanced over the average intensity behind the sample. We experimentally demonstrate focusing of light through random photonic media using both an amplitude only mode liquid crystal spatial light modulator and a MEMS-based spatial light modulator.
Our use of Micro Electro-Mechanical System (MEMS)-based digital micromirror devices for the control of the incident light field opens an avenue to high speed implementations of wavefront shaping.
\end{abstract}



\section{Introduction}
In many random photonic materials such as paper, paint and biological tissue light is multiply scattered. As a result, the propagation of light becomes diffuse and the materials appear to be opaque. Nevertheless, it has recently been demonstrated that it is possible to control light propagation through such samples by manipulating the incident wavefront \cite{Vellekoop2007, Vellekoop2008a, Vellekoop2010b, Popoff2010a, Popoff2010b, Yaqoob2008, Cui2010b, Cui2010a, Hsieh2010b}. An example for controlling light propagation by wavefront manipulation is optical phase conjugation (OPC), where a field that exits from the strongly scattering sample is phase conjugated and sent back to retrace its path to reconstruct the intensity pattern of the original incident field \cite{Yaqoob2008, Cui2010b, Cui2010a, Hsieh2010b}. OPC is successful in reconstructing a field through random photonic media, however, it does not provide a one-way focusing of light through such samples. First demonstration of one-way focusing of light through \cite{Vellekoop2007}, or inside \cite{Vellekoop2008a} strongly scattering materials was achieved by spatially modifying the phase of the incident light wave pixel by pixel using an algorithm to compensate for the disorder in the sample. It was shown that the shape of the focus obtained with this method is independent of experimental imperfections and has the same size as the speckle correlation function \cite{Vellekoop2010b}. A related approach to control light propagation by wavefront manipulation was demonstrated by Popoff and coworkers. They measured part of the optical transmission matrix, and used it to create a focus\cite{Popoff2010a} and reconstruct an image behind the strongly scattering sample \cite{Popoff2010b}. All of these methods require modulating the phase of the incident wavefront. Therefore the speed of the utilized phase modulator becomes a limiting factor on the applicability of the method to materials whose configuration change rapidly, such as biological samples \cite{Cui2010b}.

In this paper we introduce a new focusing method based on binary amplitude modulation. The wave incident to the turbid material is spatially divided into a number of segments. A portion of these segments are selectively turned off. In contrast to existing wavefront shaping methods, the phase of the segments is not modified. We demonstrate two implementations of this method to focus light through a multiply scattering TiO$_{2}$ sample; one using a liquid crystal on silicon (LC) spatial light modulator (SLM) in amplitude-only modulation mode and the other using a digital micromirror device (DMD). DMDs consist of millions of mirrors that can be independently controlled to reflect light either to a desired position or to a beam dump. This effectively switches light coming from a particular pixel of DMD on or off and provides a way to spatially modulate the amplitude of light in a binary fashion. The advantage of DMDs over LC SLMs lie in their switching speed. An important figure of merit for switching speed is the settling time, which is the time required for a pixel to become stable after changing its state. For a standard DMD the settling time is 18 $\mu$s \cite{Dudley2003}, which is approximately three orders of magnitude faster than that for typical LC SLMs used in the previous works \cite{Cui2010a, Cui2010b, Hsieh2010b, Vellekoop2007, Vellekoop2008a, Vellekoop2010b, Popoff2010a, Popoff2010b}. Such fast devices as DMDs have the potential to create a focus behind turbid material in time scales shorter than required for the configuration of the sample to change, hence can prove useful for focusing light through biological tissue \cite{Cui2010b}.

We describe the algorithm that is used to create a focus behind a turbid material by selectively turning off the segments of the SLM in Section 2. Implementation of the method using an LC SLM is described in Section 3. In this section, we present measurements of the enhancement of intensity inside the created focus and compare the results to the enhancements expected under ideal situations. In Section 4, we demonstrate focusing light through a turbid material using a MEMS-based SLM. In the Appendix, derivation of an analytical formula for the intensity enhancement from the binary amplitude modulation algorithm is provided.

\section{The binary amplitude modulation algorithm}

Light transport through a strongly scattering sample can be described using the concept of a transmission matrix that connects incident and outgoing scattering channels. Scattering channels are the angular or spatial modes of the propagating light field \cite{Beenakker1997}. In this paper, we will denote
incident and outgoing scattering channels as input and output channels, respectively. At the back of the sample the electric field of light at each output channel is related to the electric field of light at each input channel by the transmission matrix of the sample\cite{Goodman2000}:
\begin{equation}
E_{m}=\sum_{n=1}^{N}t_{mn}E_{n},
\label{eq:tm}
\end{equation}
where $E_{m}$ is the electric field at the $m^{th}$ output channel; $E_{n}$ is the electric field at the $n^{th}$ input channel; and $t_{mn}$ are the elements of the transmission matrix.

In our experiments a light beam incident to a strongly scattering sample is spatially divided into a number of square segments. Each segment corresponds to a specific range of incident angles to the sample. When input channels are described in terms of angular modes of incident light field, each SLM segment covers a range of input channels. As the SLM is divided into more segments, the angular resolution is increased and more input channels are independently controlled. We image the back surface of the sample with a CCD camera. In this case, each diffraction limited spot corresponds to an output channel. In the present experiments we select a single target output channel and use the algorithm to maximize the intensity.

\begin{figure}[H]
\centering
\includegraphics[width=12.5cm]{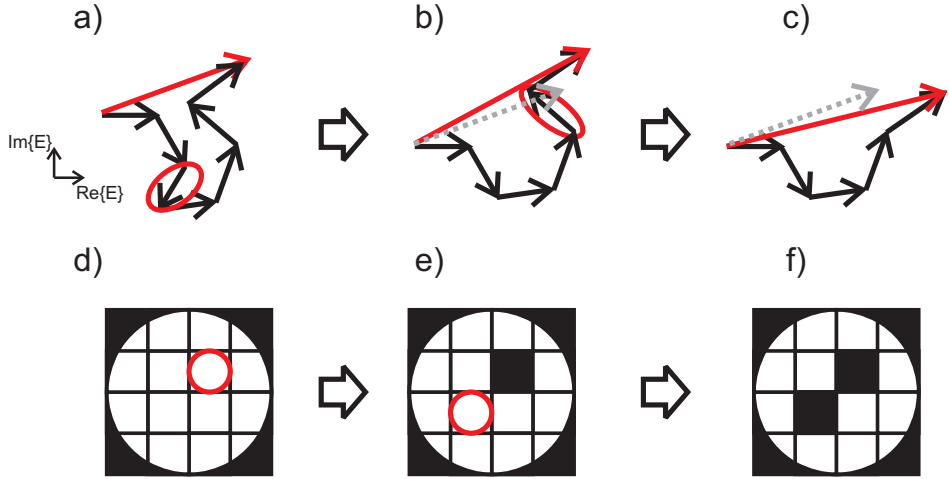}
\caption{(color) Graphical explanation of the binary amplitude modulation algorithm.
(a-c) Complex plane representation of the electric field at the target in successive steps of the algorithm. Small black vectors represent the electric field of each input channel as modified by traveling through the sample. The red vector is the total electric field at the target output channel. Dashed gray vector represents the electric field at the target position before optimization.
(d-f) Evolution of the amplitude pattern on the SLM.
(a,d) In this step, a segment which contributes negatively to the total amplitude is identified (circled). This segment will be turned off as algorithm proceeds to next segment.
(b,e) Subsequently, other segments which contribute negatively are identified and will be turned off.
(c,f) At the end of the algorithm, all of the segments which have a negative contribution to the total electric field at the target are turned off.
}
\label{fig:principle}
\end{figure}

The working principle of the algorithm is illustrated schematically in Fig. \ref{fig:principle}. In the top panel, we see a vectorial representation of the electric field in the selected target channel, $E_{m}$. This electric field is a vectorial sum of electric fields of all incident channels multiplied by the corresponding transmission matrix element. With the algorithm all segments of the incident field are successively probed. Each segment is turned on and off while the intensity at the target output channel is being monitored. This procedure can be visualized by following the block arrows in Fig. \ref{fig:principle} (a-c; d-f). As a result the segments leading to destructive interference with the resultant electric field are turned off and the intensity at the target is increased as compared to the unoptimized case. This increase can be seen by comparing the magnitudes of the red vectors in Fig. \ref{fig:principle} (a) and  Fig. \ref{fig:principle} (c). The evolution of the amplitude pattern on the SLM can be visualized by following Fig. \ref{fig:principle} (d-f).

When the algorithm is complete, a two dimensional binary amplitude pattern is obtained on SLM; by sending less light to the sample, more light is concentrated to the position of the focus. This is conceptually similar to focusing light by a conventional Fresnel zone plate \cite{BornWolf2003}. In fact, with spatial binary amplitude modulation, reconfigurable and high degree of freedom Fresnel zone plates are actively created and utilized to focus light through a strongly scattering material.

\section{Experiments with a Liquid Crystal Spatial Light Modulator}
\begin{figure}[H]
\centering
\includegraphics[width=12cm]{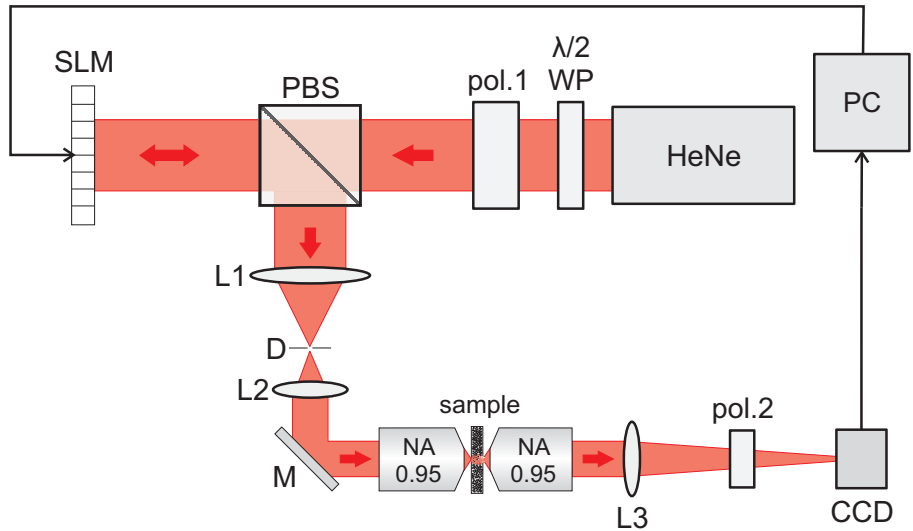}
\caption{
(color) Experimental setup. A HeNe laser beam with a wavelength of 632.8 nm and output power of 5 mW is expanded and passed through a half waveplate ($\lambda$/2 WP), a polarizer (pol.1) and a polarizing beam splitter (PBS) to be reflected off a Holoeye LC-R 2500 liquid crystal spatial light modulator (SLM). Phase and amplitude modulation is decoupled \cite{Putten2008}. A high NA (NA=0.95) microscope objective projects the shaped wavefront on the sample and an identical microscope objective collects the light transmitted through the sample. The transmitted intensity pattern is passed through a polarizer (pol.2) and monitored with a CCD camera. The computer (PC) receives intensity pattern from the CCD and adjusts the SLM segments according to the algorithm. L1, 250 mm focal length lens. D, aperture. L2, 150 mm focal length lens. M, mirror. L3, 600 mm focal length lens.
}
\label{fig:setup}
\end{figure}

The setup used in the experiments is shown in Fig. \ref{fig:setup}. A HeNe laser (wavelength 632.8 nm, output power 5 mW) is used as the light source. We pass the beam through a half waveplate and a Glan-Taylor polarizer to obtain vertically polarized light with adjustable power. The beam is expanded with a 30X beam expander (not shown) and sent to a polarizing beam splitter cube (PBS). The vertically polarized light is transmitted through the PBS to fall on the twisted nematic liquid crystal SLM (Holoeye LC-R 2500). Using the technique described in \cite{Putten2008}, we can turn a segment of the SLM on or off without changing the phase. Light is projected on to the sample by a 63X 0.95-NA infinity corrected Zeiss microscope objective. Light transmitted through the sample is collected with an identical microscope objective, passed through a polarizer and imaged on to a CCD camera with a 600 mm focal length lens, L3.  The effective magnification of the imaging system is 229X. The sample is a 36.5$\pm$3.1 $\mu$m thick layer of airbrush paint (rutile TiO$_{2}$ pigment with acrylic medium). The transport mean free path for similar samples are $l_{tr}$=0.55$\pm$0.1$\mu$m at 632.8 nm wavelength \cite{Vellekoop2007}.

\begin{figure}[H]
\centering
\includegraphics[width=10cm]{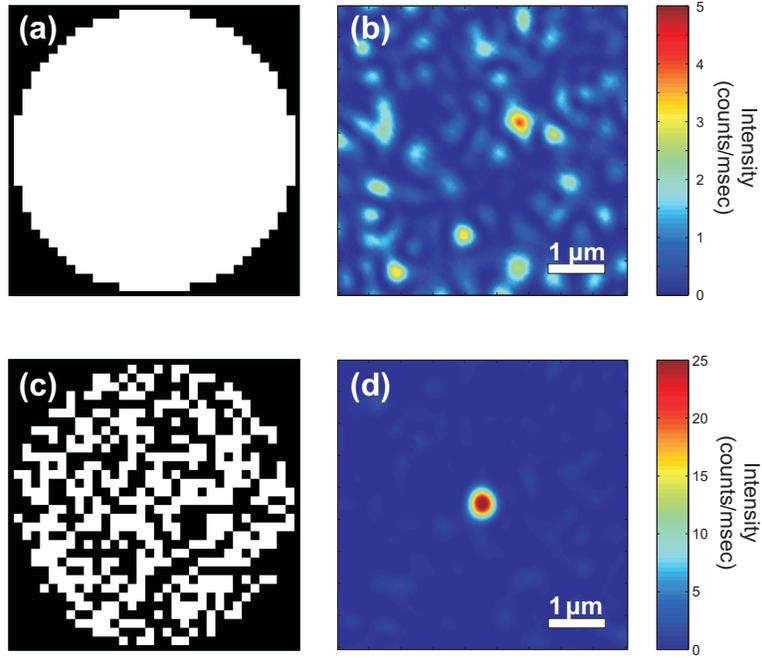}
\caption{(color) Experimental results of the optimization procedure. (a) Amplitude map written on the SLM before optimization. Active area of the SLM is divided into 812 segments, all of which are on. (b) Image captured by the CCD before optimization is performed. (c) Amplitude map on the SLM after the optimization procedure is complete. (d) Image captured after the optimization is complete. A single bright spot appears on the target point. Note the different color scale between (b) and (d).}
\label{fig:spot}
\end{figure}

The images captured with the CCD camera before and after the optimization are shown in Fig. \ref{fig:spot}, along with the amplitude map on the SLM. In this case, the SLM is divided into 812 segments. Before the optimization all segments are on and the transmitted intensity pattern is random speckle, Fig. \ref{fig:spot} (a, b). After the optimization about half of the segments are off and the transmitted intensity pattern is dominated by a single bright spot in the position of target output channel, Fig. \ref{fig:spot} (c, d). This demonstrates that using spatial binary amplitude modulation, light can be effectively focused behind a multiply scattering medium.

To have a quantitative measure of the contrast between the bright optimized spot and the background, the intensity enhancement, $\eta$ is defined as:
\begin{equation}
\eta\equiv\frac{I_{\mathrm{opt}}}{I_{\mathrm{ref}}},
\label{eq:etadef}
\end{equation}
where $I_{\mathrm{opt}}$ is the optimized intensity inside the target area after spatial binary amplitude modulation is performed for a specific sample and $I_{\mathrm{ref}}$ is the reference intensity. To measure a suitable reference intensity, the wavefront that is shaped to give a bright focus at target is sent to different parts of the sample. The intensities measured in target with changing sample configuration are ensemble averaged to give $I_{\mathrm{ref}}$. The enhancement we obtain with this definition gives a measure of the contrast between the focus and the background of the image since the reference intensity is approximately the same as the average background intensity. Since nearly half of the segments on the SLM are turned off in the optimized wavefront, the reference intensity is approximately half of the ensemble averaged intensity when all segments are on.

\begin{figure}[H]
\centering
\includegraphics[width=10cm]{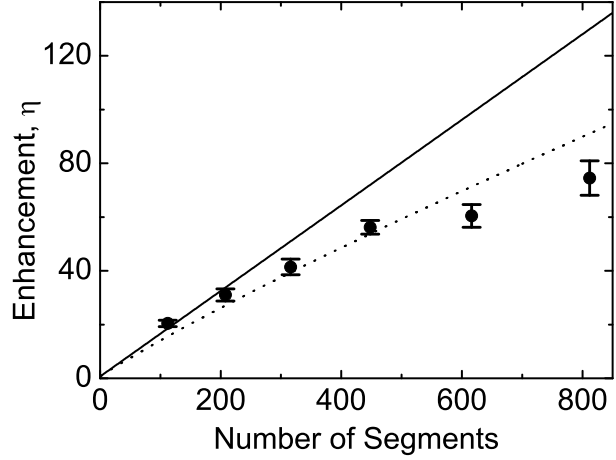}
\caption{Intensity enhancement at the target position versus the number of segments on the SLM. Black solid line: enhancements expected under ideal conditions, as obtained from Eq. (\ref{eq:etalarge1}). Each data point (black circles) is an ensemble average of 14-25 data points obtained from measurements. Bars represent the standard error of each measurement set. Black dotted curve: fit performed for the experimental enhancements using Eq. (\ref{eq:etanoise1}) with a single free parameter, SNR. Best curve fit yields SNR=24.}
\label{fig:etasimexp}
\end{figure}

In Fig. \ref{fig:etasimexp} we show the measured enhancement values as well as the ideally expected enhancement values. We measured the enhancements for a wide range of segments. The enhancement increases as the SLM is divided into more segments since the number of independently controlled input channels increases. The ideal enhancement $\eta_{\mathrm{ideal}}$ increases linearly with the number of controlled input channels N as
\begin{equation}
\langle\eta_{\mathrm{ideal}}\rangle\approx1+\frac{1}{\pi}\left(\frac{N}{2}-1\right).
\label{eq:etalarge1}
\end{equation}
However, deviations from the ideal conditions reduce the intensity enhancement. We have derived that intensity enhancement under intensity noise, $\langle\eta_{\mathrm{non-ideal}}\rangle$ can be written as
\begin{equation}
\langle\eta_{\mathrm{non-ideal}}\rangle=\langle\eta_{\mathrm{ideal}}\rangle\left(\frac{1}{2}+\frac{1}{\pi}\mathrm{arctan}\left(\frac{SNR}{\sqrt{N}}\right)\right)\frac{\langle A\rangle^{2}}{\langle A^{2}\rangle},
\label{eq:etanoise1}
\end{equation}
where SNR represents the signal to noise ratio of the signal at target position, and $\langle A\rangle^{2}/\langle A^{2}\rangle$ is a factor introduced to account for non-uniform illumination of the SLM, with A representing the amplitude of field reflected from each SLM segment. When the illumination pattern of the SLM is investigated, $\langle A\rangle^{2}/\langle A^{2}\rangle$ is found to be 0.97$\pm$0.01. Derivation of Eq. (\ref{eq:etalarge1}) and Eq. (\ref{eq:etanoise1}) can be found in the Appendix. The experimental data are fitted to Eq. (\ref{eq:etanoise1}) using the signal to noise ratio (SNR) as the only adjustable parameter. The value of the adjusted SNR is found to be 24. From a test performed on the experimental setup with a static binary amplitude pattern on the SLM, the intensity fluctuations of the light incident to the sample was measured and found to have an SNR of 165. The fact that the adjusted SNR has a lower value than measured SNR can be caused by several reasons: in the experiments the state of each segment is updated continuously during an optimization, increasing the rate of wrong decisions as the optimization proceeds. However,
Eq. (\ref{eq:etanoise1}) assumes that the probability of making a wrong decision for the state of a segment is constant throughout the optimization process. Moreover, Eq. (\ref{eq:etanoise1}) takes only intensity noise into account, which is an incomplete description of possible sources of noise or instabilities in the experimental setup. Further investigation of effects of noise and instabilities on the performance of the presented algorithm is beyond the scope of this paper.

Although the implemented algorithm was found to be sensitive to environmental factors, our experimental data convincingly shows that light can be focused through turbid materials using spatial binary amplitude modulation. In our experiments, light intensity at the target position was found to be enhanced up to 75$\pm$6 times the average speckle intensity in the background.

\section{Experiments with a Micro Electro-Mechanical System Based Spatial Light Modulator}
Spatial binary amplitude modulation enables the application of MEMS-based devices such as the digital micromirror devices in wavefront shaping experiments. In this section, we describe demonstration of focusing of light through a turbid medium using a MEMS-based SLM to modulate the wavefront. To our knowledge this demonstration is the first MEMS based focusing through turbid media. The SLM that is employed in the experiments described in this section is a disassembled projector (Sharp multimedia projector XR-32X-L), containing a digital micromirror device from Texas Instruments.

\begin{figure}[H]
\centering
\includegraphics[width=12cm]{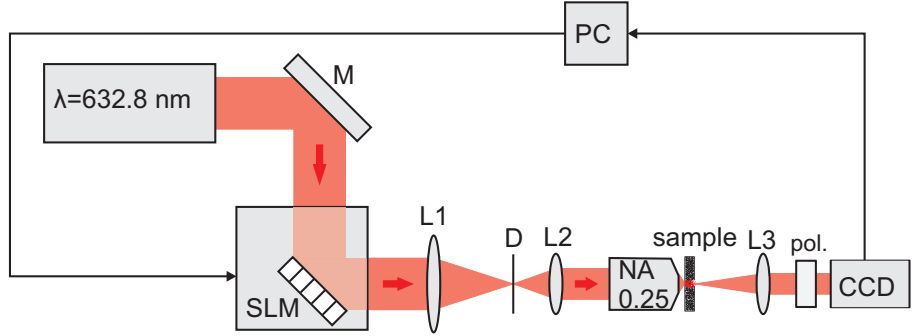}
\caption{
(color) Experimental setup for MEMS-based focusing. A HeNe laser beam that has a wavelength of 632.8 nm and output power of 2 mW is expanded and used to illuminate the SLM via a mirror (M). L1, L2 and L3 are planoconvex lenses with respectively 150 mm, 50 mm and 50 mm focal lengths. D is an aperture used for spatial filtering, and NA 0.25 is a microscope objective having 10X magnification and 0.25 numerical aperture. Light exiting the sample is converted to far field with L3, passed through a polarizer, pol. and projected on a CCD camera, which is connected to the SLM via a PC.}
\label{fig:setupMEMS}
\end{figure}

The setup used in the MEMS based focusing experiments is shown in Fig. \ref{fig:setupMEMS}. A HeNe laser, which has a wavelength of 632.8 nm and an output power of 2 mW is used as the light source. The beam is expanded with a 10X beam expander (not shown) and sent to the digital micromirror device (DMD) based SLM.  The DMD consists of 1024$\times$768 square mirrors each having a size of 10.91$\times$10.91 $\mu$m. Each mirror can exhibit two angles; it either reflects light to the intended target or into a light dump \cite{Dudley2003}. Light reflected from the DMD is projected onto the sample by a 10X 0.25 NA microscope objective and light transmitted through the sample is passed through a polarizer and projected on the CCD camera with a 50 mm focal length lens. The sample that is used in the experiments described in this section is 18.5$\pm$2.4 $\mu$m thick layer of airbrush paint (rutile TiO$_{2}$ pigment with acrylic medium). The transport mean free path for similar samples are $l_{tr}$=0.55$\pm$0.1$\mu$m at a wavelength of 632.8 nm \cite{Vellekoop2007}.

\begin{figure}[H]
\centering
\includegraphics[width=10cm]{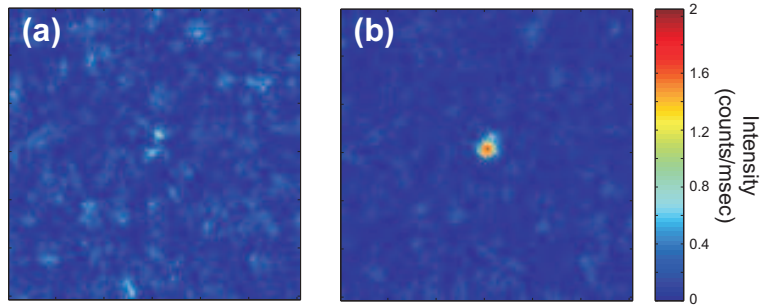}
\caption{(color) (a) Image of an area of 121 by 121 
pixels of the camera is presented just before the optimization process. (b) The same area is presented after the optimization process was finished.
In both figures, the intensity is measured in counts/milliseconds and presented on the same scale. The SLM is divided into 3228 segments.}
\label{fig:resultMEMS}
\end{figure}

The images captured with the CCD camera before and after the optimization are shown in Fig. \ref{fig:resultMEMS}. We successfully focused light through a layer of paint using a MEMS based device. The intensity enhancement is defined here as the ratio of the average intensity inside the bright optimized spot to the average intensity outside the spot, and the highest intensity enhancement that could be obtained with the setup in Fig. \ref{fig:setupMEMS} was found to be 19. However, an ideal enhancement of 514 is expected from Eq. (\ref{eq:etalarge1}). The low enhancements are thought to be the consequence of the DMD being embedded into a commercial display projector, introducing undesirable features for our purpose. Such features include turning off of the pixels of the DMD with a predefined timing, which we could not control. Lack of mechanical damping or control of noise sources in the setup is proposed to be another reason for obtaining a reduced enhancement in our experiments.

The DMDs are known to have very fast switching between on and off states and a settling time of 18 $\mu$s \cite{Dudley2003}. In our experiments, the optimizations using a DMD chip was achieved in a time scale of several minutes, which is similar to time scales of optimizations performed using the LC SLMs. This effect is due to addressing the device via the video card of the PC, which was performed in the same manner for both SLMs, limiting the communication speed to 60 Hz. With faster control of the DMD devices and use of faster cameras for detection, the speed of the method will increase close to three orders of magnitude and the method will be useful for focusing through materials whose configuration change in short time scales, like biological tissue and can be used for medical imaging purposes.

\section{Conclusion}
We have demonstrated focusing of light through strongly scattering materials by spatially modulating the amplitude of the incident field. From experiments, we have obtained an enhancement of 75 of the target intensity, when the incident wavefront is divided into 812 independently controlled segments. We have also implemented the method using a commercial projector that has a MEMS-based digital micromirror device as the spatial light modulator, providing the first demonstration of MEMS-based focusing of light through turbid materials. Use of MEMS technology will enable a fast and versatile way to control light through turbid materials.

\section{Acknowledgment}
We thank Jacopo Bertolotti, Ad Lagendijk and Pepijn Pinkse for valuable discussions; Rick van Keken for help in modification of Sharp multimedia projector. This work is part of the research
program of the ``Stichting voor Fundamenteel Onderzoek der Materie
(FOM)", which is financially supported by the ``Nederlandse
Organisatie voor Wetenschappelijk Onderzoek (NWO)". The research of APM is supported by a Vidi grant from NWO.

\appendix

\section{Analytical Expression for Ideal Intensity Enhancement}
When no optimization is performed in the system a plane wave is sent to the sample and the amplitude and phase of the electric field at each input channel is equal:
\begin{equation}
E_{n}=Ae^{i\phi}.
\end{equation}
Since the phase of incident field is assumed to be constant for an unoptimized wavefront, it can as well be taken as $\phi$=0, so that $E_{n}=A$ when no optimization is performed and $E_{n}$ is either 0 or $A$ after optimization is complete.

For a multiple scattering sample, phases of the transmission matrix elements, $\mathrm{arg}(t_{mn})$ have a uniform distribution between $-\pi$ and $\pi$ \cite{Goodman2000}. The amplitudes of the transmission matrix elements, $|t_{mn}|$, on the other hand are approximated by a Rayleigh probability density function. The electric field at the target output channel is a vectorial sum of random phasors:
\begin{equation}
E_{m}=\sum_{n=1}^{N}|t_{mn}|e^{i\mathrm{arg}(t_{mn})}E_{n}.
\end{equation}
Reference light intensity at the target position is the ensemble average of intensities recorded in the target for different sample configurations:
\begin{equation}
I_{\mathrm{ref}}=\langle E_{m}^{*}E_{m}\rangle,
\end{equation}
\begin{equation}
I_{\mathrm{ref}}=\langle\sum_{k}^{N'}A|t_{mk}|e^{-i(\mathrm{arg}(t_{mk}))}\sum_{n}^{N'}A|t_{mn}|e^{i(\mathrm{arg}(t_{mn}))}\rangle,
\end{equation}
the same wavefront is assumed to be sent to the sample while both the intensity inside the focus and the reference intensity are calculated, so that $N'$ is the number of segments that remain on after the optimization procedure is finished. It is important to emphasize that the wavefront is optimized for a certain configuration of the sample and is effectively a randomly shaped wavefront for a different configuration of the sample. So, while $I_{\mathrm{ref}}$ is calculated, light coming from different input channels have random phases at the target position. In this case we assume that the phase of each vector constituting $E_{m}$ is drawn from a distribution that is uniform between $-\pi$ and $\pi$. Using this assumption and the fact that the transmission matrix elements and the incident field are statistically independent, the reference intensity can be written as:
\begin{equation}
I_{\mathrm{ref}}=N'\langle t^{2}\rangle\langle A^{2}\rangle,
\end{equation}
where the modulus of a transmission matrix element, $t$ is a random variable having a Rayleigh probability density function.

If light fields interfere constructively at a certain position, a bright field will be obtained at that position. For this purpose, we probe the projection of the field coming from various input channels on the resultant electric field. The algorithm decides whether a segment shall be on after comparing the intensities at the target for the segment being on and segment being off cases. A segment is kept on if it contributes positively to the intensity at the target position. The contribution of the $k^{th}$ segment to the target intensity, $\Delta I_{k}$ is:
\begin{equation}
\Delta I_{k}=|E_{m}|^{2}-|E_{m}-E_{k}t_{mk}|^{2}.
\label{eq:summation}
\end{equation}
Since $\Delta I_{k}$ is a sum of uncorrelated random variables, it has a Gaussian distribution due to the Central Limit Theorem. Therefore the number of segments that remain on after the optimization is determined by the probability of drawing a positive random variable from the distribution:
\begin{equation}
f(\widetilde{\Delta I_{k}})=\frac{1}{\sigma\sqrt{2\pi}}e^{-\frac{(\widetilde{\Delta I_{k}}-\mu)^{2}}{2\sigma^{2}}},
\label{eq:idealDeltaIk}
\end{equation}
where $\widetilde{\Delta I_{k}}$ is a random variable representing $\Delta I_{k}$.
This distribution has a mean value of
\begin{equation}
\mu=A^{2}\langle t^{2}\rangle,
\label{eq:1}
\end{equation}
and a standard deviation of
\begin{equation}
\sigma=\sqrt{\langle t^{4}A^{4}\rangle +(2N-3)\langle t^{2}A^{2}\rangle^{2}}.
\label{eq:sigma}
\end{equation}
The number of segments that are on after the optimization is:
\begin{eqnarray}
N'=NP(x>0)&=&\int_{0}^{\infty}\frac{N}{\sigma\sqrt{2\pi}}e^{-\frac{(x-\mu)^{2}}{2\sigma^{2}}}dx,\\
&=&\frac{N}{2}\mathrm{erfc}\left(\frac{-\mu}{\sigma\sqrt{2}}\right).
\label{eq:correction}
\end{eqnarray}
We assume that the phases of the segments that remain on are uniformly distributed between $(-\pi/2,\pi/2)$ after the optimization, so we have:
\begin{eqnarray}
\langle I_{\mathrm{opt}}\rangle &=&\langle E_{m}^{*}E_{m}\rangle,\\
&=&N'\langle A^{2}t^{2}\rangle+N'(N'-1)\langle At\rangle^{2}\frac{4}{\pi^{2}}.
\end{eqnarray}
Under ideal conditions, i.e. when noise and instabilities are ignored, the ensemble averaged intensity enhancement at the target position, $\langle\eta_{\mathrm{ideal}}\rangle$ is found to be:
\begin{equation}
\langle\eta_{\mathrm{ideal}}\rangle=\frac{\langle I_{opt}\rangle}{\langle I_{ref}\rangle}=1+\frac{1}{\pi}(N'-1).
\label{eq:eta2}
\end{equation}
When the number of controlled input channels is large, $\langle\eta_{\mathrm{ideal}}\rangle$ becomes:
\begin{equation}
\langle\eta_{\mathrm{ideal}}\rangle\approx1+\frac{1}{\pi}\left(\frac{N}{2}-1\right).
\label{eq:etalarge}
\end{equation}
A factor of $\pi^{2}/2$ more intensity enhancement can be obtained from phase modulation \cite{Vellekoop2007}. This is expected since with phase shaping method, all $E_{n}$ are actively assigned a phase leading to total constructive interference at the target while with binary amplitude shaping active modification of the phases is not performed. The remarkable fact that the enhancement from a 1-bit modulation method can be comparable to a full analog phase modulation has been observed previously in the context of one-channel acoustic time-reversal experiments \cite{Derode1999}.

In deriving Eq. (\ref{eq:etalarge}) the amplitude of the fields in all input channels were assumed to be the same. However, in our experiments, a Gaussian beam impinges on the SLM and the amplitude of each input channel's field is modified accordingly. This introduces a prefactor of $\langle A\rangle^{2}/\langle A^{2}\rangle$ to the theoretically expected enhancement \cite{vanBeijnum2009}. In the experiments described in Section 3, this prefactor is found to have a value of 0.97$\pm$0.01.

\section{Analytical Expression for Intensity Enhancement Under Intensity Noise}
\begin{figure}[H]
\centering
\includegraphics[width=10cm]{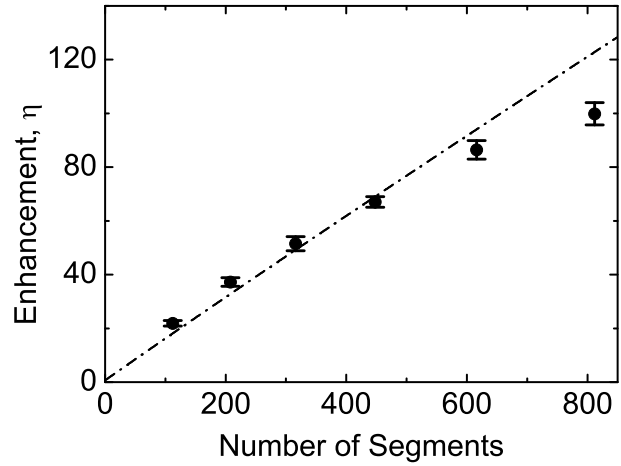}
\caption{Numerically simulated intensity enhancement at the target position versus the number of segments that the SLM is divided into. Each data point represented by the black circles is an ensemble average of a set of data points obtained from simulations conducted with an intensity noise of SNR=165. Bars represent the standard error of each measurement set. The dashed line shows the enhancements obtained from Eq. (\ref{eq:etanoise}), assuming $\langle A\rangle^{2}/\langle A^{2}\rangle=1$ and using Eq. (\ref{eq:etalarge}).}
\label{fig:etasimth}
\end{figure}

We now proceed to include the effect of noise on the intensity enhancement. We take into account noise due to intensity fluctuations of the incident light to the sample. Noise affects the correctness of the decision on whether to keep each segment of the SLM on or off. Under noisy conditions $P_{\mathrm{wrong}}$ is the probability for the algorithm to make a wrong decision for the state of a single segment, \textit{i.e.}, keeping it on while it should be turned off and vice versa. This probability is:
\begin{equation}
P_{\mathrm{wrong}}=P(\Delta I_{k}>0\bigwedge\Delta I^{exp}_{k}<0)+P(\Delta I_{k}<0\bigwedge\Delta I^{exp}_{k}>0),
\end{equation}
where $\Delta I^{exp}_{k}$ is the experimentally measured difference between the target intensities for on and off states of the $k^{th}$ segment; $P(\Delta I_{k}>0\bigwedge\Delta I^{exp}_{k}<0)$ is the probability of experimentally measuring a negative $\Delta I^{exp}_{k}$ while under ideal conditions, $\Delta I_{k}$ is positive. Likewise, $P(\Delta I_{k}<0\bigwedge\Delta I^{exp}_{k}>0)$ is the probability of experimentally measuring a positive $\Delta I^{exp}_{k}$ while under ideal conditions $\Delta I_{k}$ is negative.
\begin{equation}
P_{\mathrm{wrong}}=\int_{0}^{\infty}f(\widetilde{\Delta I_{k}})\int_{-\infty}^{0}f(\widetilde{\Delta I^{exp}_{k}})d\widetilde{\Delta I^{exp}_{k}}d\widetilde{\Delta I_{k}}+\int_{-\infty}^{0}f(\widetilde{\Delta I_{k}})\int_{0}^{\infty}f(\widetilde{\Delta I^{exp}_{k}})d\widetilde{\Delta I^{exp}_{k}}d\widetilde{\Delta I_{k}}.
\label{eq:Pwrongintegral}
\end{equation}
Here $\widetilde{\Delta I_{k}}$ is a random variable representing $\Delta I_{k}$ and has the probability density function as given in Eq. (\ref{eq:idealDeltaIk}). Similarly, $\widetilde{\Delta I^{exp}_{k}}$ is a random variable representing $\Delta I^{exp}_{k}$ and has the probability density function:
\begin{equation}
f(\widetilde{\Delta I^{exp}_{k}})=\frac{1}{(\sqrt{2})\sigma_{\mathrm{noise}}\sqrt{2\pi}}e^{-\frac{(\widetilde{\Delta I^{exp}_{k}}-\Delta I_{k})^{2}}{4\sigma_{\mathrm{noise}}^{2}}},
\label{eq:expDeltaIk}
\end{equation}
with $\widetilde{\Delta I^{exp}_{k}}$ having a mean of $\Delta I_{k}$ and a standard deviation of $\sqrt{2}\sigma_{\mathrm{noise}}$, where $\sigma_{\mathrm{noise}}$ is the standard deviation for noise. Thus, $P_{\mathrm{wrong}}$ can be evaluated as:
\begin{equation}
P_{\mathrm{wrong}}=\frac{1}{2}-\frac{1}{\pi}\mathrm{arctan}\left(\frac{\sigma}{\sqrt{2}\sigma_{\mathrm{noise}}}\right).
\label{eq:Pwrongsimplified}
\end{equation}
From Eq. (\ref{eq:sigma}), it is reasonable to assume that $\sigma=\sqrt{2N}\langle A^{2}t^{2}\rangle$ for large N. $\sigma_{\mathrm{noise}}$ can be written as $\sigma_{\mathrm{noise}}=\langle I_{m}\rangle/SNR=N\langle A^{2}t^{2}\rangle/SNR$. Substituting $\sigma$ and $\sigma_{\mathrm{noise}}$ in Eq. (\ref{eq:Pwrongsimplified}), we obtain:
\begin{equation}
P_{\mathrm{wrong}}=\frac{1}{2}-\frac{1}{\pi}\mathrm{arctan}\left(\frac{SNR}{\sqrt{N}}\right).
\label{eq:Pwrongcomplete}
\end{equation}
The possibility of making wrong decisions for a segment leads to observation of a reduced intensity enhancement as compared to the ideal case. The intensity enhancement at target position including the effects of noise and a Gaussian illumination profile, $\langle\eta_{\mathrm{non-ideal}}\rangle$ is given by:
\begin{equation}
\langle\eta_{\mathrm{non-ideal}}\rangle=\langle\eta_{\mathrm{ideal}}\rangle\left(\frac{1}{2}+\frac{1}{\pi}\mathrm{arctan}\left(\frac{SNR}{\sqrt{N}}\right)\right)\frac{\langle A\rangle^{2}}{\langle A^{2}\rangle}.
\label{eq:etanoise}
\end{equation}
In Fig. \ref{fig:etasimth} we compare the enhancements obtained from Eq. (\ref{eq:etanoise}) with the enhancements obtained from simulations for a case where incident light fluctuates with SNR=165. Other sources of noise or instabilities are neglected. It can be seen from Fig. \ref{fig:etasimth} that as N increases the enhancements obtained from the simulations become lower than the enhancements expected from Eq. (\ref{eq:etanoise}). We attribute this to the fact that each segment's state is continuously updated in the simulations (as is the case for the experiments), so that the error rate increases as the optimization progresses, which is more dominant for large N. This dynamic increase of the error probability is not taken into account in the derivation of Eq. (\ref{eq:etanoise}). Further investigation of the effects of noise on the quality of obtained foci is beyond the scope of this paper and is an interesting subject for further studies.

\end{document}